\documentclass[twocolumn,10pt]{tsfp}
\usepackage{flushend}
\usepackage{graphicx}
\usepackage[authoryear,round]{natbib}
\usepackage{fancyhdr}
\usepackage{xcolor}
\usepackage{bm}
\usepackage{hyperref}
 
\newcommand\Rm{\mbox{\textit{Rm}}} 
\newcommand\Pra{\mbox{\textit{Pr}}} 
\newcommand\Rey{\mbox{\textit{Re}}} 

\newcommand\Gr{\mbox{\textit{Gr}}} 
\newcommand\Ha{\mbox{\textit{Ha}}} 

\title{Heat transport in magnetohydrodynamic duct flow regimes with conducting and insulating walls}

\author{Andreu Queralt
    \affiliation{
	Institut für Thermo- und Fluiddynamik \\
	TU Ilmenau\\
	P.O.Box 100565, D-98684 Ilmenau\\
    andreu.queralt-mcbride@tu-ilmenau.de
    }	
}

\author{Dmitry Krasnov
          \affiliation{Institut für Thermo- und Fluiddynamik\\
	TU Ilmenau\\
	P.O.Box 100565, D-98684 Ilmenau\\
	dmitry.krasnov@tu-ilmenau.de
    }
}

\author{Yuri Kolesnikov
          \affiliation{Institut für Thermo- und Fluiddynamik\\
	TU Ilmenau\\
	P.O.Box 100565, D-98684 Ilmenau\\
	 yuri.kolesnikov@tu-ilmenau.de
    }
}

\author{Jörg Schumacher 
          \affiliation{Institut für Thermo- und Fluiddynamik\\
	TU Ilmenau\\
	P.O.Box 100565, D-98684 Ilmenau\\
	joerg.schumacher@tu-ilmenau.de
    }
}

\begin{document}

\maketitle   
\thispagestyle{fancy}

\fontsize{9}{11}\selectfont

\section*{ABSTRACT}
The flow of a liquid metal (LM) in a rectangular duct segment, subject to a uniform transverse magnetic field and uniform heating at the side walls is explored in an ample parameter space using Direct Numerical Simulation (DNS). We modify electrical wall conductivity, (either highly conducting or perfectly insulating) and investigate the effects of the buoyancy force, both in horizontally and vertically orientated ducts. In the latter case, it may be directed either with the flow or against the flow, creating backflow regions. In this parameter space and with the presence of vortex promoters at the inlet of the duct we identify $4$ types of flow. We calculate the Nusselt number $Nu(t)$ for each of them and study the statistical properties to compare their heat transfer capabilities in future fusion reactor blankets.

\section*{INTRODUCTION}
One of the main candidates for coolant in future fusion reactors are liquid metals which flow inside the blankets. They satisfy operational requirements and, due to the flow of neutrons from the plasma core, tritium breeding can occur to provide fuel for future reactions. Despite these advantages, several engineering questions regarding their viability are still open.  In the TOKAMAK context, the flow is subject to strong magnetic fields. The toroidal field component in particular is  present in the blanket. Depending on the orientation of the duct, this component will be either in the streamwise direction or along two of the side walls. A second, smaller component is present in the poloidal direction of the TOKAMAK. For simplicity we will consider only the toroidal field, which will  be perpendicular to the streamwise direction. 

In the case of conducting duct walls, the Lorenz force increases, dampens the velocity and increases the pressure drop $\Delta P$ needed to pump the metal with sufficient volume flux \citep{bib:MHD_HistoryTrends}. Additionally, an M-profile is formed at the Shercliff layers. The strong velocity gradients at the walls are detrimental, both to the production of tritium and the lifespan of the duct, with the steep velocity gradients increasing corrosion at the walls \citep{bib:corrosion}.

To overcome these problems, ceramic Flow Channel Inserts (FCI) are mounted at the duct walls. These come with their own set of problems, such as different heat dilatation from the metal, development of cracks and disintegration under the neutron flux, the latter of which increases maintenance requirements (see, e.g., \cite{Smolentsev:2021} for an overview).

\section*{DUCT GEOMETRY}
\label{sec:setup}
\begin{figure}
	\centering
	\includegraphics[width=1\linewidth]{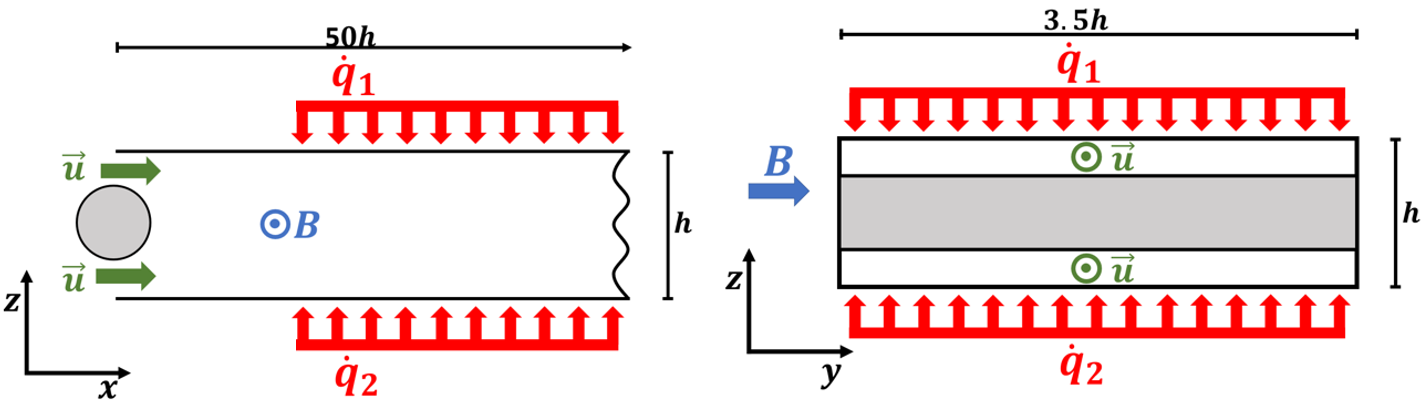}
    \caption{Simulated setup schematic for the duct. Left: side view. Right: downstream view.}
	\label{fig_setup}
    \vskip-6mm
\end{figure}

Figure \ref{fig_setup} shows the simulation domain, a rectangular duct of aspect ratio $L_y/h = 3.5$ is considered. To maintain cohesion with previous studies, such as in \cite{Pamm:2023}, the length of the duct $L_x$ is $50$ times its height $h$ and at the inlet we also prescribe a velocity profile with two flat jets near the top and bottom walls. This profile mimics the velocity field past a round cylinder, which was used in experiments \citep{Pamm:2023} to generate inlet instabilities by covering $\sim 50\%$ of the inlet area (see figure \ref{fig_setup}).
When heating is applied it is done only at the Shercliff walls, while the Hartmann walls remain adiabatic.  Both Shercliff walls are heated to simplify the problem by imposing symmetry. The heating area begins at $x=20$, again to maintain cohesion with previous studies. When the conductivity of the wall is present, it is set equal at all four walls. 

\section*{MATHEMATICAL MODEL}
\label{sec:matmodel}
In order to solve the flow of the liquid metal, the governing equations are used for an incompressible flow at low magnetic Reynolds numbers $\Rm$, known as quasi-static approximation (see, e.g., \cite{Davidson:2016}):
\begin{eqnarray}
\label{eq:NS}
    \nabla \cdot \bm{u} &=& 0
    \\
    \frac{\partial\bm{u}}{\partial t}\! + \!(\bm{u} \!\cdot\! \nabla)\bm{u} &=& \!-\nabla\! p + \!\frac{1}{\Rey}\!\bigg(\!\nabla^2\!\bm{u}+\Ha^2\!\bm{j}\!\times\!\bm{e}_B\!\bigg)\! - \!\frac{\Gr}{\Rey^2}T\!\bm{e}_g
    \\
    \frac{\partial T}{\partial t}+\bm{u}\cdot \nabla T &=& \frac{1}{\Rey \Pra}\nabla^2T
    \\
    \bm{j} &=& -\nabla \phi + \bm{u}\times \bm{e}_B
    \\
    \nabla^2\phi &=& \nabla\cdot(\bm{u}\times \bm{e}_B)
\end{eqnarray}

Here $\bm{u}$, $p$, $T$, $\bm{j}$ and $\phi$ are the velocity, pressure, temperature, current density and electric potential fields. For conducting cases, we use the thin wall approximation (see \cite{bib:in_house_solver} and references therein). This model assumes that the currents close within the (thin) walls. Non-dimensional parameters $\Rey$, $\Ha$, $\Pra$ and $\Gr$ are the Reynolds, Hartmann, Prandtl and Grashof numbers respectively, defined as $\Rey = U_0 L/\nu$,  $\Ha = B L (\sigma/\rho\nu)^{1/2}$, $\Pra = \nu k^{-1}$, $\Gr = g \beta \Delta T L^3 \nu^{-2}$,  where  $\Delta T$ is the temperature deviation from a reference temperature. The half-height of the duct $L = h/2$ and the mean velocity $U_0$ in the streamwise direction are used as the length and velocity scales. Finally, $\bm{e}_g$ and $\bm{e}_B$ are the unit vectors in the directions of acceleration due to gravity and the imposed magnetic field. \newline
The equations are solved numerically using our in-house finite-difference solver described in (\cite{bib:in_house_solver}).

\section*{THE NUSSELT NUMBER}

\begin{figure}
    \centering
    \includegraphics[width=\linewidth]{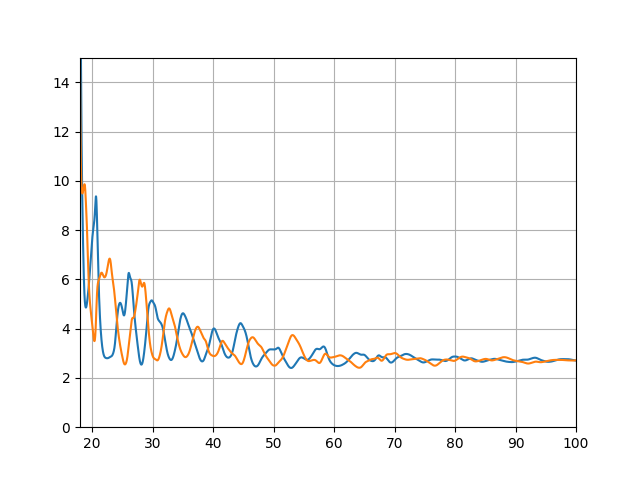}
    \caption{$Nu(x)$ at the top and bottom walls for $Re=4000$, $Ha=325$ $c_W=0$ and passive heat transport}
    \label{fig:NuExample}
\end{figure}

To compare the heat transfer efficiency of the different flow regimes, we will use the Nusselt number $Nu$, defined as \citep{bib:ConvectionHeatTransfer}:
\begin{equation}
    Nu(x,t)=\frac{q'L}{k(T_w(x,t)-T_b(x,t))}
\end{equation}
Where $q'$is the heat flux across the duct wall, $L$ the duct half-height, $k$ the thermal conductivity, $T_b$ the bulk temperature and $T_w$ the wall temperature. For our simulations, we computed $Nu(x,t)$ using both the top and bottom walls and found that the results were similar, for the insulating duct runs the most notable difference was with a $180$ phase shift between them. For the conducting duct, the phase difference was not constant. Figure \ref{fig:NuExample} shows an example of $Nu(x,t)$ where this is clearly visible.
The average $Nu(x,t)$ along the duct length is calculated with equation \ref{eq:Nu}. Where $l_0$ is the starting position for heating and $l_f$ is the duct length. Due to the imposed symmetry of the problem, this value is statistically the same when measured from both walls.
\begin{equation}
    Nu(t)=\int^{l_f}_{l_0}Nu(x,t)dx
    \label{eq:Nu}
\end{equation}
For the remainder of the paper, we take $Nu(t)$ to be the duct-averaged $Nu(t)$ at the top wall ($z=+h/2$). In addition, $\overline{Nu}$  will be the time average of $Nu(t)$:
\begin{equation}
    \overline{Nu} = \int_{t_0}^{t_f}Nu(t)dt,
\end{equation}
where $t_0$ is taken once the flow is fully developed. For our case, we take $t_0=200$ convective time units, which is the longest initial transient we encountered (see $QM$ flow in figure \ref{fig:1000TKE}). The end point of our simulations is $t_f=600$ in convective units. 

\section*{SIMULATION PARAMETERS}
Table \ref{tab:parameters} shows the parameters used for the simulations performed for this paper. We also performed simulations with $c_W = 0.1$ with upward and downward flows and did not see any noticeable effect. The simulations were also performed at $Ha=500, 750$, however, the results are not shown in this paper. To evaluate the efficiency of the vortex promoters, we also performed simulations with laminar flow at the inlet, but only for the duct in a horizontal position. All simulations were run on a $4096 \times 320 \times 92$ grid with uniform meshing in the streamwise $x$-direction and strong clustering towards the walls in the $y$- and $z$-directions.

\begin{table*}
    \centering
    \begin{tabular}{cccccccl}
         Simulation&  $Re$&  $Ha$&  $Pr$&  $Gr$&  $c_W$ &Flow Direction &Observed Flow\\ \hline
         1&  $4000$&  $325$&  $0.02$&  $10^6$&  $0.0$&Horizontal &$QH$\\
         2&  $4000$&  $325$&  $0.02$&  $10^6$&  $0.1$&Horizontal &$UL$\\
         3&  $4000$&  $325$&  $0.02$&  $10^6$&  $0.0$&Upwards &$QM$\\
         4&  $4000$&  $325$&  $0.02$&  $10^6$&  $0.0$&Downwards &$QW$\\
 5& $4000$& $1000$& $0.02$& $10^6$& $0.0$& Horizontal &$QH$\\
 6& $4000$& $1000$& $0.02$& $10^6$& $0.1$& Horizontal &$UL$\\
 7& $4000$& $1000$& $0.02$& $10^6$& $0.0$& Upwards &$QM$\\
 8& $4000$& $1000$& $0.02$& $10^6$& $0.0$& Downwards &$QW$\\ \hline
    \end{tabular}
    \caption{Simulation parameters and observed flow type.}
    \label{tab:parameters}  
\end{table*}

\section*{THE FOUR FLOW TYPES }
\label{sec:flowTypes}
\begin{figure*}
    \centering
    \includegraphics[width=1\linewidth]{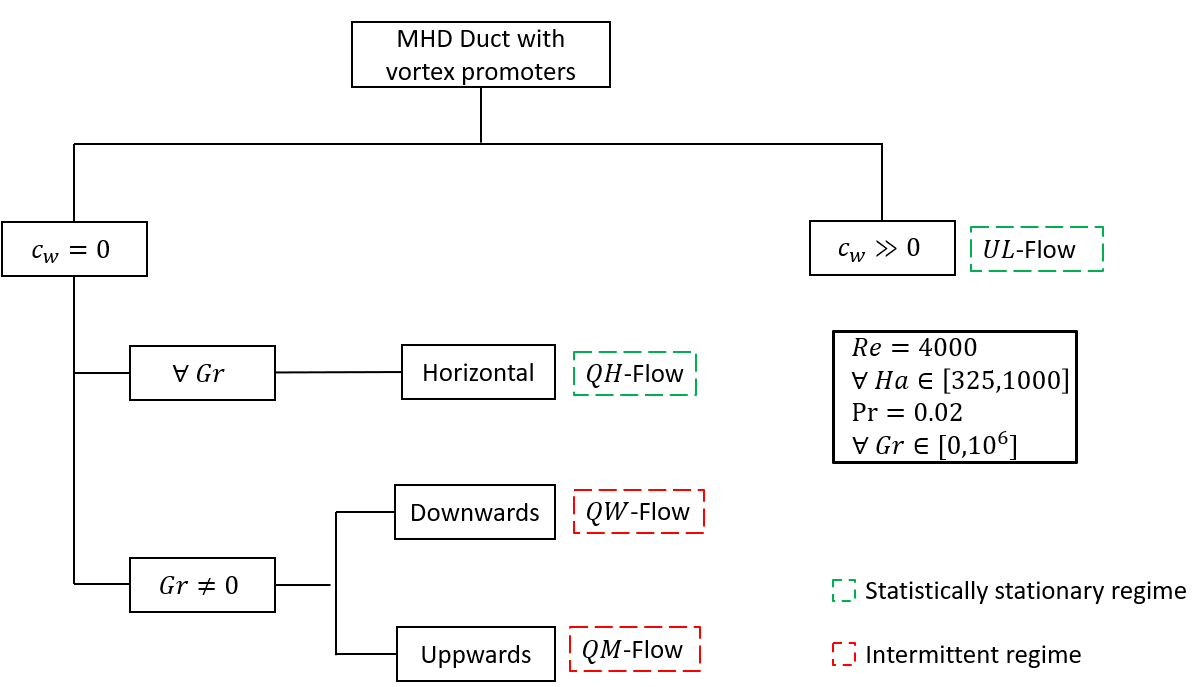}
    \caption{Schematic overview of the parameters at which each flow type exist.}
    \label{fig:parameterSchematic}
\end{figure*}

We classify the flow regimes into four categories. Figure \ref{fig:parameterSchematic} shows the parameters at which each flow type has been encountered.

\begin{enumerate}
    \item $UL$-Flow: Statistically stationary Unstable  Walker flow with $M$-Profile between the Shercliff walls.
    \item $QH$-Flow: Statistically stationary quasi-2D ($Q2D$) rolls. These rolls eventually decay into a Hartmann flow.
    \item $QW$-Flow: Intermittent $Q2D$ rolls with backflow jets ($W$ shape) at the Shercliff layers.
    \item $QM$-Flow: Intermittent $Q2D$ rolls with side jets at the Shercliff layers ($M$ shape, but weaker than in the $UL$ flow)
\end{enumerate}
The first type of flow occurs when the duct walls are highly conducting. All other types of flow occur with perfectly insulating ducts. The intermittent regimes occur when the duct is in a vertical position. In those situations, the buoyancy force is large enough to  modify the velocity profile at the Shercliff layers, where the current is perpendicular to the magnetic field and thus the Lorenz force is zero. At the bulk and Hartman layers, the buoyancy force acts to modify the velocity magnitude, but since the Lorenz force is proportional to it, the effects are neutralized. This creates either an $M$-shaped (when the mean flow is directed upwards) or a $W$-shaped profile (when the mean flow is directed downwards) at the side walls. Eventually velocity gradients become large enough to trigger instabilities, which eventually die out and the original state is restored.

For the case of perfectly insulating walls with $c_W=0$ at higher $Ha$, there is a lengthy transient before the flow reaches its final state. Initially, two vortex streaks appear instead of one. In both vertical duct orientations, these streaks eventually become just one. In the horizontal duct, the flow linearizes, and after some time develops a single streak. 

In highly conducting ducts, the Shercliff sidewall jets develop instabilities known as jet-detachments, often found even at low-$Re$ regimes \citep{BraidenEPL:2016}. These detachments can be triggered either by the inlet vortex promoters or naturally, due to the strong velocity gradients with inflection points \citep{Priede:2010,Arlt:2017}, formed between the Shercliff layers. This, once again, is observed at the higher values of $Ha$. 

\begin{figure*}
    \includegraphics[width=1\linewidth]{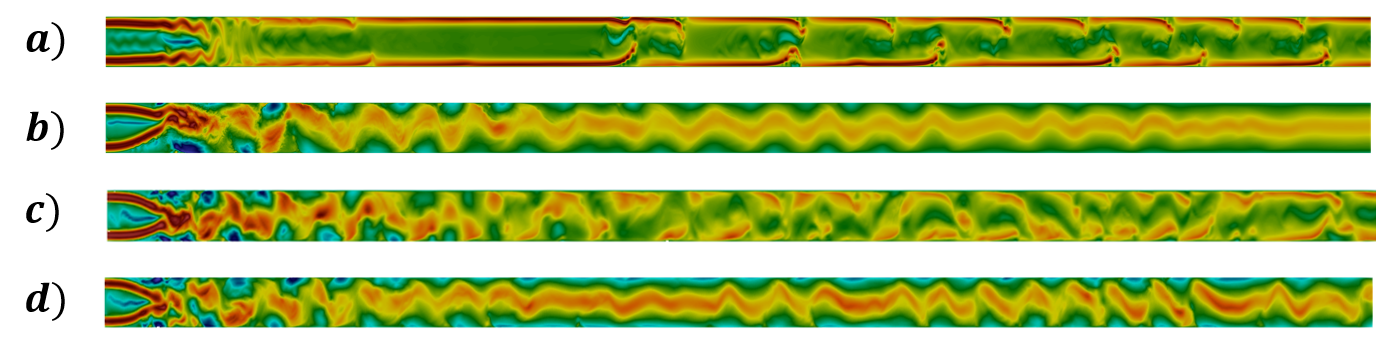}
    \caption{ $M-$profile with detachments in the Shercliff layers in a conducting duct$a)$, $Q2D$ rolls in a perfectly insulating horizontal duct $b)$, upwards-flow duct with buoyancy driven side wall jets $c)$ and downwards-flow duct with buoyancy driven backflow side wall jets $d)$ All four are at $Re=4000$, $Ha=325$, $Gr=10^6$ and $Pr=0.02$. $c_W=0.1$. at  $a)$ and  $c_W=0.0$ at $b), c)$ and $d)$ }
    \label{fig:flowSnapshots}
\end{figure*}

In this study, we focus on results at $Ha=325$ and $1000$, which are the lowest and highest values that we have considered. Figure \ref{fig:flowSnapshots} shows instantaneous snapshots for each of the flow types at $Ha=325$.
\section*{RESULTS}
\begin{figure}
    \centering
    \includegraphics[width=1\linewidth]{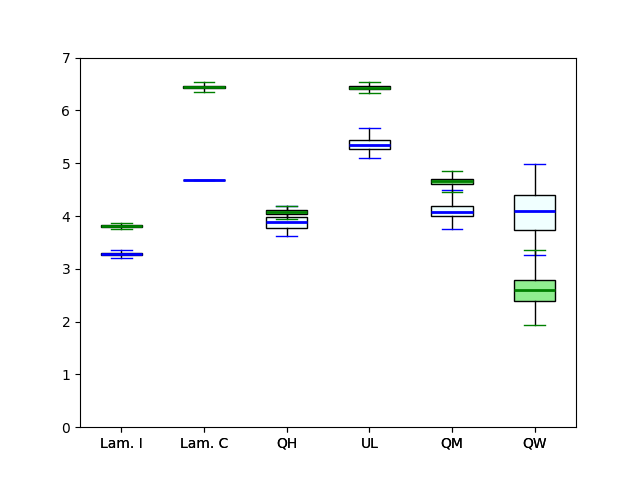}
    \caption{Box plots of $\overline{Nu}$ without vortex promoters with insulating walls (Laminar I), conducting walls (Laminar C) and each of the four regimes encountered with vortex promoters at $Ha=325$ (blue) and $Ha=1000$. Outliers have been remove from the plot to avoid cluttering.}
    \label{fig:NuBoxPlot}
\end{figure}

Table \ref{tab:NuValsPromoters} shows the time averaged $Nu$ numbers and their standard deviation for all four regimes at $Ha=325$ and $Ha=1000$ in the presence of vortex promoters. Table \ref{tab:NuValsLaminar} shows the same for the horizontal flow in the absence of vortex promoters. Together with figure \ref{fig:NuBoxPlot} they provide a birds eye view of the dependence of $\overline{Nu}$ and its time fluctuations with $Ha$ and the flow type. In general, $\overline{Nu}$ Increases with $Ha$, with the exception of the $QW$ floe, in which $Nu$ decreases significantly. This regime also has the highest variance.  In both cases, the highest $Nu$ was observed by the $UL$ flow.  At $Ha=1000$ the vortex promoters have no effect, since the disturbances occur naturally. At $Ha=325$, however, the flow is laminar without the promoters. Despite this, the $Nu$ is still higher than for all three insulating wall cases at $Ha=325$, which are very similar to each other. This is because in the $UL$ flow as a fluid parcel is heated by conduction from the walls, it is quickly advected downstream by the high-speed sidewall jets. This results in lower fluid temperatures close to the walls, which reduces the temperature gradient between the walls and the bulk, resulting in higher $Nu$. At higher $Ha$, the side jets of the $QM$ flow become more pronounced, so this flow type ends with a better $Nu$ than the $QH$ and $QW$ flows.

\begin{table}
    \centering
    \begin{tabular}{ccccc}
 & \multicolumn{2}{c}{$Ha=325$} & \multicolumn{2}{c}{$Ha=1000$}\\ \hline
         Flow Type&  $\overline{Nu}$& std($\overline{Nu}$) & $\overline{Nu}$&std($\overline{Nu}$) \\ \hline
         $Q$&  $3.88$& $0.138$& $4.07$&$0.0508$\\
         $UL$&  $5.36$& $0.138$& $6.43$&$0.041$\\
         $QM$&  $4.09$& $0.152$& $4.65$&$0.090$\\
         $QW$&  $4.08$& $0.411$& $2.61$&$0.303$\\ \hline
    \end{tabular}
    \caption{ Time-averaged $Nu$ for the entire duct length and their standard deviation for $Re=4000$, $Gr=10^6$, $Pr=0.02$ at $Ha=325$ and $Ha=1000$. Simulations with vortex promoters}
    \label{tab:NuValsPromoters}
\end{table}
\begin{table}
    \centering
    \begin{tabular}{ccccc}
 & \multicolumn{2}{c}{$Ha=325$} & \multicolumn{2}{c}{$Ha=1000$}\\ \hline
         $c_W$&  $\overline{Nu}$& std($\overline{Nu}$) & $\overline{Nu}$&std($\overline{Nu}$) \\ \hline
         $0.0$&  $3.27$& $0.0465$& $3.80$&$0.0434$\\
         $0.1$&  $4.68$& $3.65\cdot10^{-5}$& $6.44$&$0.356$\\ \hline
    \end{tabular}
    \caption{ Time-averaged $Nu$ for the entire duct length. And their standard deviation for $Re=4000$, $Gr=10^6$, $Pr=0.02$ at $Ha=325$ and $Ha=1000$. Base case without vortex promoters}
    \label{tab:NuValsLaminar}
\end{table}

\begin{figure}
    \centering
    \includegraphics[width=1\linewidth]{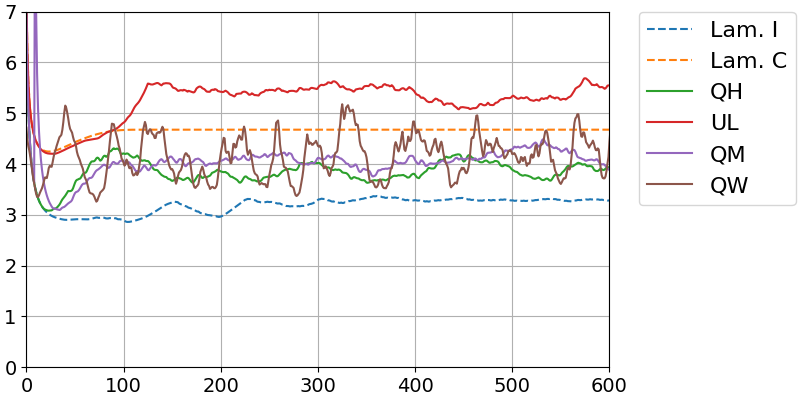}
    \caption{Temporal evolution of $\overline{Nu}$ for the four flow types over $600$ convective time units at $Ha=325$, including initial transient. Dashed lines represent simulations without vortex promoters}
    \label{fig:325Nu(t)}
\end{figure}

\begin{figure}
    \centering
    \includegraphics[width=1\linewidth]{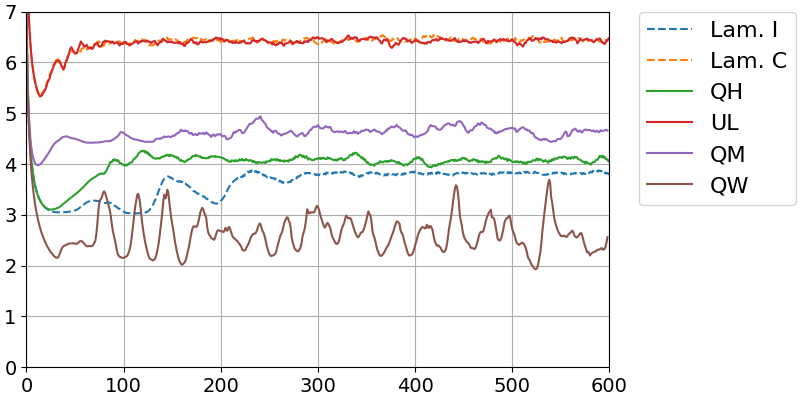}
    \caption{Temporal evolution of $\overline{Nu}$ for the four flow types over $600$ convective time units at $Ha=1000$, including initial transient}
    \label{fig:1000Nu(t)}
\end{figure}

\begin{figure}
    \centering
    \includegraphics[width=1\linewidth]{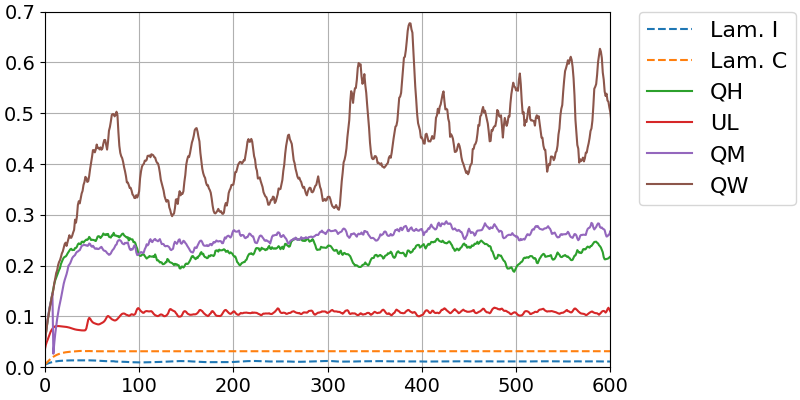}
    \caption{Temporal evolution of the volume averaged duct TKE for 600 convective time units at $Ha=325$, including initial transient. Dashed lines represent simulations without vortex promoters}
    \label{fig:325TKE}
\end{figure}

\begin{figure}
    \centering
    \includegraphics[width=1\linewidth]{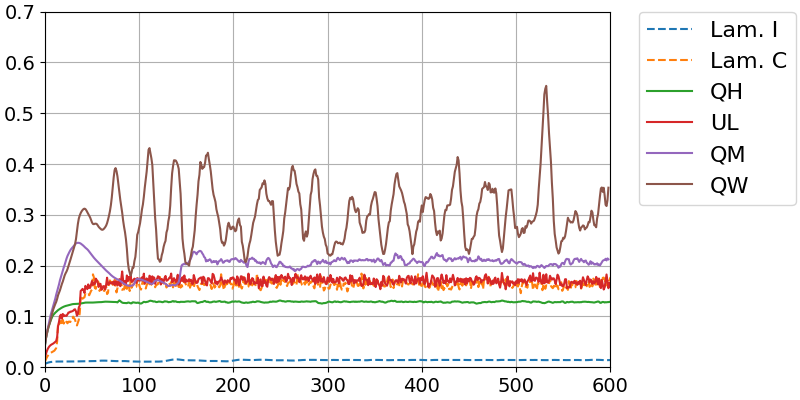}
    \caption{Temporal evolution of the volume averaged duct TKE for 600 convective time units at $Ha=1000$, including initial transient}
    \label{fig:1000TKE}
\end{figure}

Figures \ref{fig:325Nu(t)} and \ref{fig:1000Nu(t)} show the time evolution of $\overline{Nu}$, and figures \ref{fig:325TKE} and \ref{fig:1000TKE} show the $TKE$ of the same flows. The $QW$ flow, which has poor $Nu$-performance has the highest $TKE$, followed by the $QM$ flow, indicating that those are better at mixing. This is because the same fluid parcel that was quickly advected out of the duct in the $UL$ flow is advected not only in the stream wise direction but also strongly in the vertical direction. This causes the heat from the neutron flux to remain longer in the duct, and therefore higher temperatures, specially close to the wall and a lower $\overline{Nu}$
\section*{DISCUSSION}
The results presented here lead to a seemingly non-straightforward conclusion. Namely, for the parameter space that we considered, it turns out that good heat transfer regimes are poor mixing regimes and vice-versa.

\begin{figure}
    \centering
    \includegraphics[width=0.5\linewidth]{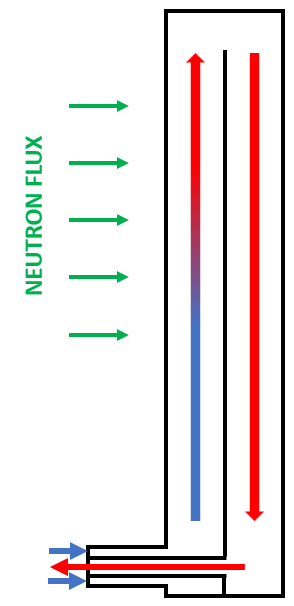}
    \caption{Possible blanket configuration with fluid in the neutron-facing wall being pumped upwards and the flowing back downwards, ensuring best possible mixing.}
    \label{fig:blanket}
\end{figure}

\begin{figure}
    \centering
    \includegraphics[width=0.8\linewidth]{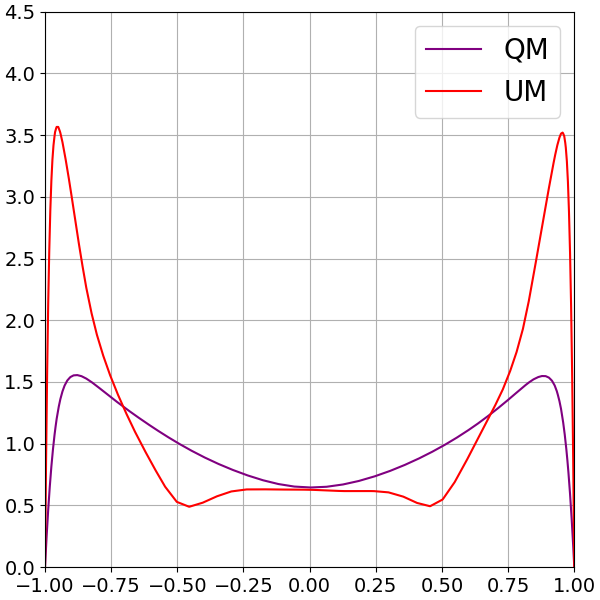}
    \caption{Time averaged velocity over 200 convective time units between the Shercliff layers at $x=100$ fir the $UL$ and $QM$ flow regimes}
    \label{fig:sideJets}
\end{figure}

\begin{figure}
    \centering
    \includegraphics[width=0.8\linewidth]{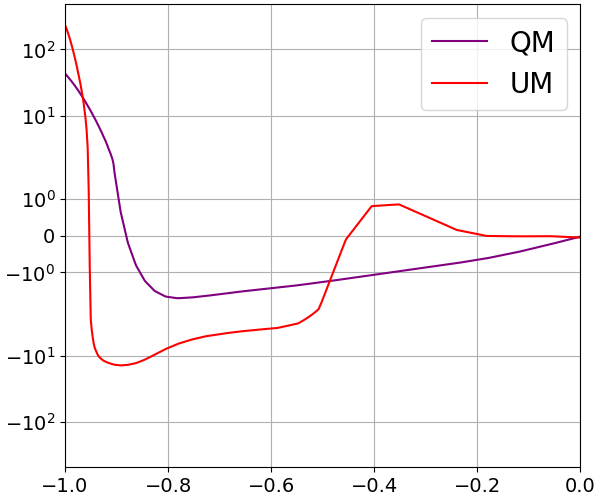}
    \caption{Time averaged velocity gradient $\partial_zu$ over 200 convective time units between the Shercliff layers at $x=100$ fir the $UL$ and $QM$ flow regimes}
    \label{fig:sideGradients}
\end{figure}

From a purely thermodynamic point of view, the best configuration for the $LM$ blanket is the conducting duct, regardless of orientation. However, the same side jets that provide optimal heat transfer are also responsible for the high velocity gradients that can cause corrosion and poor neutron breeding, as well as an unacceptable $MHD$ pressure drop.  An alternative, is to leverage the $QM$ flow, which also has side jets, albeit much weaker ones. Figure \ref{fig:sideJets} shows the magnitude of the jets of the $UL$ and $QM$ flows, and figure \ref{fig:sideGradients} their gradients in symmetric logarithmic scale. The strongest gradient is one order of magnitude smaller in the $QM$ flow than it is in the $UL$ flow, while the reduction of $Nu$ is only of about $30\%$. Figure \ref{fig:blanket} shows a proposal for how the $LM$ could be pumped in a blanket. If the pumping direction is upwards, then side jets will be present regardless of the wall conductivity. Then, as it returns downwards the liquid metal is fully mixed.

The matter of making the duct walls fully insulating however also leads to engineering problems \citep{Smolentsev:2021}. That is, the neutron flux is known to destroy the FCI used to ensure the ducts are insulating. We have conducted preliminary simulations which suggest that by insulating only the Hartmann walls, which are not exposed to the neutron flux, the resulting flow is the same as for the fully insulating case. This is because more electric current lines are allowed to close inside the duct, compared to insulating all four walls, which appears to be sufficient to reduce the net Lorenz force and the associated MHD pressure drop.
\section*{ACKNOWLEDGMENTS}
This work is supported by the Deutsche Forschungsgemeinschaft (DFG) with grants KR 4445/5-1 and SCHU 1410/36-1.
The authors also acknowledge the Gauss Center for Supercomputing e.V. (\url{https://www.gauss-centre.eu}) for providing computing time on the GCS Supercomputer SuperMUC-NG at Leibniz Rechenzentrum (\url{https://www.lrz.de}) within the computing project "pn67la".

\bibliographystyle{tsfp}
\bibliography{tsfp}

@STRING(JFM="J. Fluid Mech.")

@article{bib:in_house_solver,
  title = {{Tensor-product-Thomas elliptic solver for liquid-metal magnetohydrodynamics}},
  author = {Krasnov, D. and Akhtari, A. and Zikanov, O. and Schumacher, J.},
  journal = {J. Comp. Phys.},
  volume = {474},
  pages = {111784},
  year = {2023},
  publisher = {Elsevier}
}

@INPROCEEDINGS{Pamm:2023,
  AUTHOR =       {Krasnov, D. and Listratov, Ya. and Belyaev, I. and Kolesnikov, Yu. and Sviridov, E. and Zikanov, O.},
  TITLE =        {{MHD flow of submerged jets behind the inlet disturbance}},
  BOOKTITLE =    {{PAMM Proc. Appl. Math. Mech.}},
  YEAR =         {2023},
  url = {http://doi.org/10.1002/pamm.202200200},
  editor =       "",
  volume =       "22",
  number =       "1",
  series =       "",
  pages  =       "",
  address =      "",
  month =        "",
  organization = "",
  publisher =    {Wiley-VCH Verlag GmbH \& Co. KGaA, Weinheim},
  note =         {Special Issue: $92^{nd}$ Annual Meeting of the International Association of Applied Mathematics and Mechanics (GAMM)}
}

@book{Davidson:2016,
  author={Davidson, P. A.},
  title={Introduction to magnetohydrodynamics},
  year={2016},
  publisher={Cambridge University Press}
}

@BOOK{bib:MHD_HistoryTrends,
  AUTHOR =       {Molokov, S. and Moreau, R. and Moffat, H. K.},
  TITLE =        {{Magnetohydrodynamics: Historical Evolution and Trends}},
  PUBLISHER =    {Springer},
  YEAR =         {2007},
  address =      ""
}

@article{bib:corrosion,
    Author = {Smolentsev, S. and Saedi, S. and Malang, S. and Abdou, M.},
	Journal = {Journal of Nuclear Materials},
	Pages = {294-304},
	Title = {{Numerical study of corrosion of ferritic/martensitic steels in the flowing PbLi with and without a magnetic field}},
	Volume = {771},
	Year = {2013}
}

@BOOK{bib:ConvectionHeatTransfer,
  AUTHOR =       {A. Bejan},
  TITLE =        {{Convection heat transfer, third edition}},
  PUBLISHER =    {John Wiley \& Sons},
  YEAR =         {2004},
  address =      ""
}

@article{BraidenEPL:2016,
  author = {Braiden, L. and Krasnov, D. and Molokov, S. and Boeck, T. and B\"uhler, L.},
  title = {{Transition to turbulence in Hunt's flow in a moderate magnetic field}},
  url = {http://dx.doi.org/10.1209/0295-5075/115/44002},
  journal = {Europhys. Lett.},
  number = {4},
  volume = {115},
  pages = {084501},
  year = {2016}
}

@ARTICLE{Priede:2010,
  AUTHOR =       {Priede, J. and Aleksandrova, S. and Molokov, S.},
  TITLE =        {{Linear stability of Hunt's flow}},
  JOURNAL =      JFM,
  YEAR =         "2010",
  volume =       "649",
  number =       "",
  pages =        "115--134"
}

@ARTICLE{Arlt:2017,
  AUTHOR =       {Arlt, T. and Priede, J. and B\"uhler, L.},
  TITLE =        {{The effect of finite-conductivity Hartmann walls on the linear stability of Hunt’s flow}},
  JOURNAL =      JFM,
  YEAR =         "2017",
  volume =       "822",
  number =       "",
  pages =        "880--891"
}

@article{Smolentsev:2021,
  author = {Smolentsev, S.},
  title = {{Physical Background, Computations and Practical Issues of the Magnetohydrodynamic Pressure Drop in a Fusion Liquid Metal Blanket}},
  journal = {Fluids},
  number = {6},
  pages = {110},
  volume = {3},
  year = {2021},
  url = {https://doi.org/10.3390/fluids6030110}
}
%
%
%
%
%


\end{document}